\def\isarxiv{1}

\documentclass[journal]{IEEEtran} %

\IEEEoverridecommandlockouts                              %


\usepackage{tabularx,calc}
\usepackage{color}
\usepackage{tubscolors}

\ifx \isarxiv \undefined
	\usepackage{tikz}
	\usetikzlibrary{%
		shapes,
		automata,
		arrows,
		matrix,
		backgrounds,
		fit,
		patterns,
		decorations.markings, 
		svg.path, 
		shapes.multipart, 
		external, 
		shadows, 
		positioning, 
		calc, 
		decorations, 
		intersections, 
		angles, 
		quotes,
		decorations.pathmorphing, 
		arrows.meta, 
		patterns, 
		decorations.pathreplacing, 
		external,
		snakes,
		spy%
	}   		 	 
	\usepackage{tkz-euclide}
	\usetkzobj{all}
	\usepackage{pgfplots}
	\usepackage{pgfplotstable}
	\usepgfplotslibrary{groupplots}
	\usepgfplotslibrary{statistics}
	\usepackage{tikz-3dplot}
	\usepackage{tikz-uml}
	\AtBeginDocument{\providecommand\tikzexternaldisable{\relax}} 
	\AtBeginDocument{\providecommand\tikzexternalenable{\relax}} 
\else
	\usepackage{tikzexternal}
\fi

\usepackage[main=american,british,german]{babel}
\usepackage{siunitx}
\usepackage[%
			bibstyle=ieee,
			citestyle=numeric-comp,
			backend=bibtex,
			minnames=1,
			maxcitenames=2,
			maxbibnames=99,
			doi=true,
			isbn=false,
			url=false,
			natbib=true]
			{biblatex} 
\usepackage[caption=false, font=footnotesize]{subfig}
\usepackage{csquotes}

\ifx \isarxiv \undefined
 	\usepackage[hidelinks]{hyperref}
\else 	
	\usepackage{hyperref}
\fi

\usepackage{ifthen}
\usepackage{lipsum}

\usepackage{xstring}
\usepackage[abspath]{currfile}

\usepackage{bm} 		%

\usepackage{amssymb} 	%
\usepackage{amsmath}
\usepackage{siunitx}
\usepackage{graphicx}
\usepackage{bm} 		%
\usepackage{booktabs} 	%
\usepackage{multirow}   %
\usepackage{booktabs} 	%
\usepackage{colortbl}
\usepackage{array}

\usepackage{microtype}

\newcolumntype{P}[1]{>{\centering\arraybackslash}p{#1}}
\input{macros}

\definecolor{cr}{RGB}{102,180,211}
\definecolor{mn}{RGB}{255,127,0}
\definecolor{rs}{RGB}{190,0,80}

\ifx \isarxiv \undefined
	\input{shapes.tex}
	\input{tikzstyles.tex}
\fi

\begin{document}
	\title{Supporting Safe Decision Making Through Holistic System-Level Representations \& Monitoring -- \\ A Summary and Taxonomy of Self-Representation Concepts for Automated Vehicles}%
	\author{Marcus Nolte,
			Inga Jatzkowski,
			Susanne Ernst, and 
			Markus Maurer%
		\thanks{This  research  has partially been conducted within the  project  “UNICAR\emph{agil}”  (grant number  16EMO0285) funded by the Federal Ministry of Education and Research of Germany (Bundesministerium für Bildung und Forschung,
			BMBF).}%
		\thanks{Fundamental research toward self-representation has also been funded by the Deutsche Forschungsgemeinschaft (DFG) under the research group grant FOR 1800 "Controlling Concurrent Change" (CCC, project duration 2013-2019).}%
		\thanks{Parts of the research related to decision making have been funded by the Daimler and Benz Foundation in the project "Value Based Decision Making". (\textit {Corresponding author: M. Nolte})}%
		\thanks{The authors are with the Institute of Control Engineering, Technische Universität Braunschweig, Hans-Sommer-Str. 66, 38106 Braunschweig, Germany (e-mail: \{nolte, jatzkowsi, ernst, maurer\}@ifr.ing.tu-bs.de)}}%
	
	\markboth{%
	}%
	{}
	
	\copyrightnotice%
	\maketitle

	\begin{abstract}
		The market introduction of automated vehicles has motivated intense research efforts into the safety of automated vehicle systems.
Unlike driver assistance systems, SAE Level 3+ systems are not only responsible for executing (parts of) the dynamic driving task (DDT) \cite{sae2018}, but also for monitoring the automation system's performance at all times.
Key components to fulfill these surveillance tasks are system monitors which can assess the system's performance at runtime, e.g. to activate fallback modules in case of partial system failures.
In order to implement reasonable monitoring strategies for an automated vehicle, holistic system-level approaches are required, which make use of sophisticated internal system models.
In this paper we present definitions and an according taxonomy, subsuming such models as a vehicle's \emph{self-representation} and highlight the terms' roles in a \emph{scene} and \emph{situation} representation.

Holistic system-level monitoring does not only provide the possibility to use monitors for the activation of fallbacks.
In this paper we argue, why holistic system-level monitoring is a crucial step towards higher levels of automation, and give an example how it also enables the system to react to performance loss at a tactical level by providing input for decision making.

	\end{abstract}

	\begin{IEEEkeywords}
		Self-representation, self-perception, self-awareness, scene, situation, automated vehicle monitoring, automated vehicle safety
	\end{IEEEkeywords}

	\copyrightnotice

	\section{Introduction}
\label{sec:intro}

\IEEEPARstart{A}{utomated} vehicle technology has seen promising progress over the last years.
Many challenges in perception, planning and control have been tackled for more than three decades.
The results have been key enablers for impressive public demonstrations. 
With increasing interest in a commercial market introduction of such systems, safety for Level 3+ \cite{sae2018} systems has become a highly active field of research. 
The safety-related challenges created by relieving the driver from perception, control and surveillance tasks in highly uncertain scenarios are huge, but crucial to solve.

A key aspect of safety engineering for Level 3+ system is the question of how to ensure safe vehicle operation at system runtime.
On the one hand, this includes guaranteeing safe nominal behavior e.g. by verifying the safety of system decisions (e.g. as targeted by Intel’s Responsibility Sensitive Safety Framework (RSS) \cite{shalev-shwartz2017}) and system actions \cite{pek2018}). 
On the other hand, monitoring tasks are shifted from the driver to the system for Level 3+ systems.
Hence, monitoring mechanisms are required which enable the detection, isolation, and compensation of faults, which enable performance assessment and degradation detection for the automated driving function, and which eventually allow the vehicle to take actions to enter a risk-minimal state in case of functional degradation. 

Classic functional safety design aims at hardening systems against error propagation beyond a defined system boundary at a hardware and software level.
By providing sufficient redundancy and monitors, fall-back implementations or \nobreakdash components can be activated to continue system operation at full (fail-operational) or degraded (fail-degraded) functionality.
For the safe behavior of automated vehicles fail-operational requirements imply that errors must not propagate beyond the boundary of the vehicle automation system as a whole.

In this context, system-level monitoring has recently become a more active field of research \cite{carre2019b, colwell2018}.
Structures and models which are used for the implementation of system-level monitors are, however, greatly varying.

Existing system-level monitoring approaches range from systems relying on redundant processing paths \cite{Torngren2018} to more holistic monitoring approaches, which aim at capability-conforming system behavior \cite{maurer2000a, pellkofer2002, reschka2015f}.
A more recent holistic concept in this respect is self-awareness for automated vehicles \cite{schlatow2017f, carre2019b} which strongly relies on the explicit representation of knowledge about the system itself.

In previous publications, the models which are used to represent knowledge about the system and its capabilities have been summarized under the terms self- or ego-representation \cite{maurer2000b, maurer2000a, reschka2015f, ulbrich2015}. 
While the term self-representation is also used in robotics \cite{knoll2000a}, a concrete definition in the context of automated driving is currently not available.

At the same time, the explicit runtime representation of knowledge about the system's capabilities is more than just a conceptual contribution to the monitoring problem described above.
In our view, the explicit representation of knowledge about the operational limits of the vehicle and its current performance are key enabling concepts for safe and adaptive decision making under uncertainty.

To illustrate these concepts, this paper gives an overview about current approaches for system-level monitoring of automated vehicles (\autoref{sec:literature}).
We use this context to give a concrete definition of the term self-representation for automated vehicle systems (\autoref{sec:self-representation}).
Finally, we describe the impact of self-representation on decision-making (\autoref{sec:dec_making}) and provide an illustrative example scenario to demonstrate how a holistic self-representation for automated vehicles can assist safe decision making in \autoref{sec:example}.
	\section{Related Work}
\label{sec:literature}

The following summary of related work provides context for the definitions in \autoref{sec:self-representation}.
For this purpose, related work will cover two aspects:
On the one hand, it provides an overview about recent contributions to holistic monitoring concepts for automated vehicle systems.
On the other hand, it cites relevant terminology which will be used to provide a concise definition of the terms related to self-representation.

\subsection{Holistic Monitoring for Automated Vehicles}
The need for monitoring concepts for automated vehicles has been recognized in a variety of publications.
Those approaches can roughly be divided into two categories: 
On the one hand, there are more traditional \emph{diagnosis}-like approaches which monitor singular points in the system and resort to boolean statements if system components are working as expected \cite{stahl2020online}.
On the other hand, there are approaches, which aim at system-level or \emph{functional} monitoring \cite{schlatow2017f}.
I.e. these approaches, do not solely rely on local monitoring of singular points in the system, but rather aim at inferring additional qualitative statements how the performance of individual system parts impacts the performance of the system's overall functionality.
The latter approaches may, of course, make use of the former approaches.
Hence, they put the data which is gathered by different monitors into a larger context.
A key role for establishing what we call context here is the application of internal models of the system.
In the following, we focus on system-level monitoring approaches for automated vehicles, as the concepts contribute to our understanding of \emph{self-representation} more significantly than isolated diagnosis-like approaches.

\subsubsection{Capability-Based Approaches}
Holistic monitoring requirements for automated vehicles \cite[p. 1]{dickmanns1987a} and autonomous control systems have already been formulated more than three decades ago and autonomous control systems \cite{passino1993e} (regarding cyber pysical systems, the latter concepts have recently been "rediscovered", e.g. by \cite{sifakis2019a}).

\citet{maurer2000b} picks up these ideas and formulates the need for the ego-vehicle to have "[...] a model of itself in order to react responsibly" \cite[p. 587]{maurer2000b}.
Besides possessing knowledge of its shape (geometric model) and its dynamics (vehicle dynamics model), the vehicle needs to be aware of its capabilities in order to successfully execute its mission \cite{maurer2000a}.
In this context, the monitoring of such capabilities is mainly reflected in terms of monitoring control quality metrics to supervise the correct execution of control algorithms.

These fundamental concepts are further developed by \cite{Siedersberger2003} and \cite{pellkofer2002} who introduce the concept of ability networks as an input to a central decision making module to select appropriate behavior for an encountered situation.
In all three approaches, \emph{abilities} are seen as dedicated components responsible for executing a particular task in the system.

\citet{schroder2007} apply behavior networks, modeled after the ability networks presented by \cite{pellkofer2002} and \cite{Siedersberger2003}.
These networks are a basis for decision making and use capabilities as means to modeling if an action (e.g. a lane change) is executable.

Compared to \cite{Siedersberger2003, pellkofer2002}, \citet{bergmiller2014a} uses a slightly altered terminology, subsuming the aforementioned abilities as skills\footnote{Although skills and abilities have distinct definitions in psychology, we subsume abilities and skills under the more general term capabilities for the remainder of this paper.}.
He extends the concepts of \citet{Siedersberger2003} and \citet{pellkofer2002} and proposes to abstract the system's skills from a concrete implementation, while assigning each skill with a performance value, determined by fuzzy inference.

This idea has been developed further \cite{reschka2015f} to represent the vehicle's dynamic driving task in ability and skill graphs, capturing the dependencies among capabilities and assigning quality metrics to each node in such a graph.
Most recently, we have applied those graphs for structuring requirements during the concept phase of an ISO~26262-compliant development process for automated vehicle systems \cite{nolte2017g}.
The structured process is aimed at a traceable derivation of quality metrics for capability representation.
Furthermore, we have established skill graphs as a viewpoint in an architecture framework \cite{bagschik2018a} for autonomous vehicles.
In this context, a capability decomposition serves as an additional architecture view which contributes to the traceability of behavioral safety requirements to technical requirements during system design.

The idea of explicitly formulating necessary capabilities for the safe operation of an automated vehicle has also been formulated in a white paper \cite{wood2019} published by a consortium of several OEMs and a number of tier suppliers joined by apollo and Baidu.
These implementation-agnostic capabilities are separated into fail-safe and fail-degraded capabilities.
The authors state that, in case of a system failure, fail-degraded capabilities must be performed with a certain performance level, until a minimal risk condition is reached.
Note that the formulation chosen in \cite{wood2019} is very similar to the approach presented in \cite{nolte2017g}:
While we chose to define abstract capabilities and assign safety goals to them, capabilities as formulated in \cite{wood2019} can be seen as abstract safety goals for the vehicle.
As we suggested in \cite{bagschik2018a}, \cite{wood2019} also propose to allocate capabilities to functional blocks in the system.
Further, \cite{wood2019} suggest to allocate requirements to the elements which realize the abstract capabilities, similar to our requirement decomposition approach published in \cite{nolte2017g}.

Finally, \citet{koopman2019} list sets of system limitations that must be represented for fault management in automated vehicle systems.
They formulate those limitations in terms of required capabilities \cite{koopman2019}, while \citet{koopman2019a} also generally propose occasional "self-characterization" maneuvers for checking remaining capabilities.

\subsubsection{MAPE-K and Self-Awareness}
Recent contributions to systematic monitoring approaches in the field of automated driving stem from the so-called \emph{self-x} concepts.
Self-x concepts have been established in IBM's Autonomic Computing framework \cite{kephart2003} for large-scale computing applications and comprise self-configuration, \mbox{-optimization}, \mbox{-healing} and \mbox{-protection}.
In this context, self-awareness, i.e. the capability of a system to represent knowledge of itself and its environment \cite[p. XXV]{lewis2016}, has been described as "the fundamental basis for any self-x feature" \cite[p. 164]{kramer2011}.

Self-awareness has also been described by NASA as a basis for autonomously operating aircraft \cite{gregory2016}.
They define a "[...] self-aware aircraft, space-craft or system [...]" \mbox{\cite[p. 1]{gregory2016}}, amongst other factors, as "[...] one that is aware of its internal state, has situational awareness of its environment, [and] can assess its \emph{capabilities} currently and project them into the future [...]" \cite[p. 1]{gregory2016}.

In addition to the general concept of self-awareness, a key architectural pattern in self-x-based autonomic computing systems is the so-called MAPE-K loop \cite{kephart2003}.
MAPE-K is a knowledge-based (hence -K), \textbf{M}onitor, \textbf{A}nalyze, \textbf{P}lan, \textbf{E}xecute scheme, which puts particular emphasis on monitoring.
Monitoring in the context of Autonomic Computing is understood as monitoring an element (e.g. a hardware resource) and its external environment at the same time.
Hence it subsumes internal and external aspects for the monitoring task.

With these aspects, capability-based monitoring approaches can be seen as an important contribution to a self-aware automated vehicle.
Self-awareness is a much broader concept, establishing a semantic context for system aspects by using different kinds of models \cite{schlatow2017f}.
We will further elaborate on the relation of self-awareness and self-representation in \autoref{sec:self-representation}.

There are a number of recent publications which can be attributed to the concept of self-awareness for automated vehicles.
Some of those publications rather use the term self-awareness, while focusing on smaller monitoring aspects, e.g. for anomaly detection in single signals \cite{zaal2019a} or specific functions, such as in localization algorithms \cite{ravanbakhsh2020}.
In contrast to these contained approaches, we have introduced the concept of self-aware automated vehicles \cite{schlatow2017f} and present a general argument for cross-layer models, i.e. an explicit model-based representation of different architectural viewpoints to provide knowledge about the system structure which can be used for system monitoring.
The ideas of that publication are further developed for the architecture framework presented in \cite{bagschik2018a}.

Furthermore, the adaptation of MAPE\nobreakdash-K-like structures to automated vehicle systems has e.g. been proposed by \cite{carre2019b} and \cite{Torngren2018}:
Partially motivated by propositions made in \cite{reschka2015f, schlatow2017f, nolte2017g}, \citet{carre2019b} tailors a MAPE-K-loop toward runtime safety assurance for automated vehicles.
Following \citet{kephart2003}, the loop resides in parallel to those components of the vehicle automation system, which implement the actual automated driving functionality.
The individual elements of the loop are implemented as micro services and individual loops can be managed by superimposed MAPE-K loops.
All loops in the hierarchy contribute to and draw from shared knowledge bases.
Required knowledge is subsumed under a configuration\nobreakdash-, a capability\nobreakdash-, a goal\nobreakdash- \& adaptation\nobreakdash-, a plan\nobreakdash- and a context model represented in a number of ontologies.
The application of the framework is illustrated in a simple simulated scenario, only affecting longitudinal control of the vehicle.

The approach presented by \citet{Torngren2018} does not make this clear distinction between the managed element and the MAPE\nobreakdash-K loop.
The system consists of three MAPE\nobreakdash-K loops in a nominal and a supervisor channel.
The nominal channel includes the nominal automated driving functions to implement a sense-plan-act scheme.
The supervisor channel includes two MAPE\nobreakdash-K loops, one of which monitors the nominal channel and one which is a degraded version of the nominal channel.
Knowledge is represented in the form of safety constraints in the knowledge base of the supervisor channel and provide an explicit analysis module, which is responsible for switching between nominal and degraded functionality \cite{Torngren2018}.

While the framework is well argued, the limitation to a nominal and a supervisor channel resembles classic solutions for functional safety.
With the proposed architecture it only seems possible to switch between nominal and degraded functionality.
The paper does not mention any possibility to influence vehicle decisions in the nominal channel by the information generated in the supervisor channel, which makes tactical reactions to failures at least difficult.

\subsubsection{Other Recent Approaches}
\label{sec:other}
Other recent publications also suggest explicit monitoring of Operational Design Domains (ODDs) \cite{colwell2018} or the definition of system-level "factors" to trigger graceful degradation in case of malfunctioning system parts \cite{ishigooka2019}.
Conceptwise, this is closely related to representing safety constraints in a monitoring element, as proposed by \cite{Torngren2018}.

\citet{ishigooka2019} present a variety of architectural and design considerations for fault-tolerant vehicle systems.
While they assume that monitoring is available and that faults are detectable, no details of the monitoring architecture are given.

\citet{colwell2018} take our approaches \cite{reschka2015f, schlatow2017f} as a motivation and propose the concept of Restricted Operational Domains (RODs) \cite{colwell2018}.
The framework presents a holistic approach, trying to establish traceability between functional requirements determined in the development process and a resulting ODD.
At runtime, the system is divided into a layer which implements the automated driving function (\emph{system layer}) and a layer for supervision.
In the supervision layer, \emph{system health monitors} provide input to an \emph{ROD Manager}, which determines current ODD restrictions based on monitoring data.
An \emph{ROD Monitor} determines whether the system is operating within specification.
The monitoring results are propagated to a \emph{System Supervisor}, which selects operational modes (e.g. emergency maneuvers).

While the paper addresses a variety of interesting aspects and follows traceability as paradigm for the implementation of its monitors, further architectural considerations, such as the integration of monitoring results into the algorithms of the \emph{system layer} are considered as future work.

\subsubsection{Conclusion}
Considering the approaches described above, monitoring is often seen as an additional "layer" of the system rather than as an integral part of the runtime system, as has been argued e.g. in \cite{maurer2000a, pellkofer2002, schlatow2017f} or \cite{reschka2015f}.
From the more recent publications, only \cite{carre2019b} and \cite{colwell2018} explicitly consider this option:
\cite{carre2019b} even gives a very short simulative example of influencing runtime behavior of the system, without having an explicit "fallback layer".
In our view, an automated vehicle must be provided with both options.
On the one hand, it should be able to adapt its tactical decisions to its remaining capabilities (e.g. to restrict its behavior to ROD limitations).
On the other hand, as tactical decisions can imply time delays of seconds, critical errors must be compensated at much higher frequency.
For this reason, a redundant fallback layer could be required.

	\section{Self-Representation and related definitions}
\label{sec:self-representation}

In existing (technical) literature\footnote{Terms such as "the self" or "the ego", of course, have a strong history and associated meanings in (cognitive and behavioral) psychology. For now, we will focus on technical disciplines using those terms.}, the terms self-perception and  self-representation for technical systems have been used and described before to refer to a system generating knowledge about itself.
However, a concise definition has not been given so far, as we will argue in the following.

\subsection{Robotics}
Over the last 20 years there has been a vast number of publications dealing with aspects of self-perception and self-representation in the robotics community (cf. \cite{knoll2000a, sturm2008, lanillos2017}).

While it would be beyond the scope of this paper to include all details discussed in the robotics domain, \citeauthor{knoll2000a} give descriptions summarizing some of the most relevant aspects of self-perception and -representation \cite{knoll2000a}.
They formulate \emph{self-perception} and \emph{self-representation} as key properties of a robot to generate knowledge about itself.
Without either of these properties, according to \cite{knoll2000a}, a robot is unable to differentiate between itself and the environment and in consequence unable to assess how it impacts its environment.
For \emph{self-perception}, they differentiate between introspective sensors measuring the robot's internal states\footnote{In the following, as argued by \cite{maurer2000b}, we understand \emph{state} in a broader, i.e. more semantic, sense than in its purely control-theoretic definition regarding the value of a system's state vector.} and exteroceptive sensors capturing the robot's state in its environment.

A central aspect of \emph{self-representation} is the use of models \cite{knoll2000a}.
This at least includes models of the robot's shape.
An obvious reason for this is that otherwise, the robot would not be able to move in its configuration space without colliding with obstacles.
However, models of the robot's dynamics (i.e. the differential equations to predict its motions), rule bases, and more can be applied to enable the robot to simulate its own behavior before taking actions \cite{knoll2000a}.
While \citeauthor{knoll2000a} \cite{knoll2000a} provide a description of what is required for a robot's self-perception and self-representation, they do not give a clear definition of the terms.

\citeauthor{lanillos2017} \cite{lanillos2017} give a concrete definition for self-perception.
According to them, 

\begin{quote}
	\small
	\textit{%
		"Artificial self-perception is the machine ability to perceive its own body, i.e., the mastery of modal and intermodal contingencies of performing an action with a specific sensors/actuators body configuration."} 
	
	\cite[p. 101]{lanillos2017}
\end{quote}

The second part of this definition can be understood such that the robot needs to interpret its actions with respect to the state of its environment -- i.e. self-perception is a key property for the robot to differentiate between actions which manipulate external objects and actions which manipulate its own body.

\subsection{Automated Driving}
Elements for self- (or ego- \cite{maurer2000b}) representation in the context of automated vehicles have been described by \cite{maurer2000b}.
In line with the argumentation of \citeauthor{knoll2000a} \cite{knoll2000a}, key elements are models of the vehicle's own shape and its dynamics.
In addition, \cite{maurer2000b} includes aspects, such as the representation of the runtime system in form of running processes, hardware resources, or a model of the vehicle's communication network.
However, as in \cite{knoll2000a}, a concrete definition of the term ego-representation is not given.
The same is true of our own previous publication \cite{reschka2015f}, which cites \cite{knoll2000a} for the general concepts of the terms.

\citeauthor{reschka2017} \cite{reschka2017} provides a definition of the term self-representation for automated vehicles, which translates as follows:
\begin{quote}
	\small
	\textit{%
		"Self-representation describes an image of the system's performance in the current situation. 
		It combines the internal system state with the current situation."} 
		
		\cite[p. 144, translated from German]{reschka2017}
\end{quote}

In the authors' view, this definition is problematic:
On the one hand, it focuses solely on the performance of the system, while \cite{maurer2000b} and \cite{knoll2000a} include a number of additional aspects (or models), e.g. a robot's or vehicle's shape when describing self-representation.
On the other hand, it does not comply with the following definitions of the terms \emph{scene} and \emph{situation}, as defined in \cite{ulbrich2015}\footnote{The definitions provided in \cite{ulbrich2015} are the basis for the definitions used in ISO/PAS 21448 SOTIF \cite{iso19}.}:

Here, self-representation has been defined to be part of the \emph{scene}.
For the following discussions, we use the definition of the term \emph{scene} accordingly and will elaborate on aspects of its definition in \autoref{sec:subdefinition}:

\begin{quote}
	\small
	\textit{%
		"A scene describes a snapshot of the environment including the scenery and dynamic elements, as well as all actors’ and observers’ self-representations, and the relationships among those entities. [...]"} \cite[p. 983]{ulbrich2015}
\end{quote}

In order to relate the concept of \emph{self-representation} to decision making problems, we will also refer to the \emph{situation} as defined by \cite{ulbrich2015}:

\begin{quote}
	\small
	\textit{%
		"A situation is the entirety of circumstances, which are to be considered for the selection of an appropriate behavior pattern at a particular point of time. 
		It entails all relevant conditions, options and determinants for behavior. 
		A situation is derived from the scene by an information selection and augmentation process based on transient (e.g.	mission-specific) as well as permanent goals and values. [...]"} \cite[p. 958]{ulbrich2015}
\end{quote}

Furthermore, \autoref{sec:example} will describe a short \emph{scenario} to illustrate the concept of self-representation in more detail.
For this, we refer to a scenario as "[...] the temporal development between several scenes in a sequence of scenes." \cite[p. 986]{ulbrich2015}.

\begin{figure*}
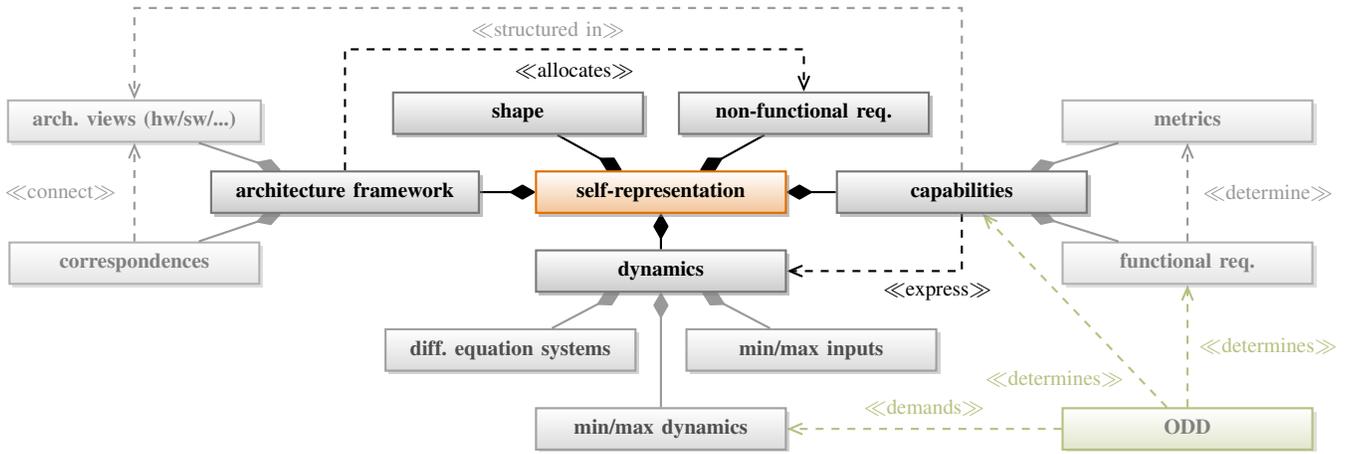

	\centering
	\includeimg{self_representation.tikz}
	\caption{Block description diagram visualizing models which are part of the system's self representation (adapted and refined from \cite{maurer2000a}). Arrows have UML semantics. Diamond arrow heads depict compositions, dashed arrows are depicting dependencies between models. Lighter blocks depict possible examples. For simplification, only the most important relations are drawn. Based on \cite[p. 59]{maurer2000a}, extended and adjusted for the differentiation between \emph{scene} and \emph{situation}.}
	\vspace{-1em}
	\label{fig:bdd}
\end{figure*}
\subsection{A Definition of Self-Representation for Automated Vehicles}
\label{sec:subdefinition}
As summarized above, a clear definition of the terms \emph{self-representation} and \emph{self-perception} for automated vehicles which conforms to the definition of \emph{scene} and \emph{situation} has not been given so far.

\subsubsection{Self-Representation}
With respect to \emph{self-representation}, the strong reference to models of the robot's (or vehicle's) internal structures as argued by \cite{maurer2000b} and \cite{knoll2000a} are much in line with the understanding of (mental) representations in cognitive psychology.
In this respect, representations are seen as "information-bearing structures (representation) of one kind or another" \cite{pitt2020}.

Hence we follow the descriptions of \cite{maurer2000a} and \cite{knoll2000a} and propose the following definition:
\begin{quote}
	\small
	\textit{%
		A system's self-representation consists of the set of explicit internal models of the system's properties. This set of models allows the system to infer knowledge about its own logic and dynamic state and to assess its own possible actions.}
\end{quote}

An overview about the models which are part of the system's self-representation is given in the block definition diagram in \autoref{fig:bdd}, which is based on \cite[p. 59]{maurer2000a}.
Compared to \cite{maurer2000a}, some of the aspects attributed to the \emph{Ego-Subject} have been subsumed under more general terms or have been separated according to the definitions of \emph{scene} and \emph{situation}.

The vehicle's \emph{shape} can subsume detailed representations such as a scene-tree \cite{maurer2000b}, or simpler descriptions such as a simple storage of the vehicle's wheel base, track width, etc. without the explicit relations provided by a scene tree.

The term \emph{architecture framework} includes explicit (cross-layer) models of the vehicle's different architectures and correspondence relations between different architectural views (e.g. as described in \cite{schlatow2017f} or \cite{bagschik2018a}). 
This e.g. includes hardware-, software- or logic/functional architectures.

Models for the vehicle's \emph{dynamics} subsume representations of the differential equations describing the vehicle's motion, the vehicle's minimum and maximum dynamics, and the physical restrictions, including by-design and fault-induced restrictions, on system inputs (e.g. min.~/~max. acceleration or steering angles).
Populating these models with sensible parameters might require identification maneuvers (\cite[p. 52]{maurer2000a}, \cite{koopman2019a}).

Models for the vehicle's \emph{capabilities} include performance metrics for each capability as well as representations of functional or technical requirements which are the basis for the associated performance metrics (cf. \cite{nolte2017g}).
The dependencies between capabilities (e.g. formulated in a graph model \cite{reschka2015f}) are structured in a dedicated architectural view of the system's architecture framework model.

Finally, we assume an explicit representation of non-functional requirements (e.g. timing requirements for controllers), which need to be monitored at runtime.
Possible relations between functional and non-functional properties of the system can again be established using the representations of the architecture framework, e.g. by performing dependency analysis across multiple architecture models (c.f. \cite{mostl2018c}, \cite{mostl2019a}).

While the targeted ODD (or \emph{domain} in \cite{maurer2000a}) has massive implications for the models in the system's self-representation, we do not consider the ODD itself as a part of the system's self-representation.
This would, in our view, diminish the role of the ODD:
The ODD impacts all parts of the system's design, as its definition process is the origin for most development assumptions.
Hence we regard the ODD as a closely related but separate concept for knowledge-representation, which governs many aspects of the vehicle's self- and environment representation (cf. \autoref{fig:bdd} depicted in green).
For further elaboration, \autoref{sec:dec_making} will provide a review of our understanding of the ODD, which conforms to the ideas presented by \cite{koopman2019, koopman2019a}.

Regarding \cite{maurer2000a}, we consider optimization criteria and the desired degree of automation to be components of the vehicle's transient goals.
The vehicle's goals have been argued to be part of the \emph{situation} representation \cite[p. 986]{ulbrich2015}, as they are more external stimuli to decision making than internals of the vehicle's self-representation. Note that this might be debatable for Level 5 vehicles, if one would like to make a case for autonomous vehicles which are allowed to choose their own mission objectives.

Aspects of the system's architectures (hardware, processes, communication \cite{maurer2000a}) have been subsumed under the representation of the architecture framework.

Finally, the vehicle's relations to other objects are part of the scene representation.

\subsubsection{Self-Perception}
Following the definition of self-representation, we define self-perception in a broader sense. 
\citet{lanillos2017}'s definition is, in the authors' opinion, too restrictive, as it is only related to the robot's body.
Conforming with \cite{knoll2000a, reschka2015f} and \cite{ulbrich2017a}, for us, self-perception rather describes a process of acquiring general knowledge about the internal state of the automated vehicle.
Just as environment perception provides an outer perspective of the vehicle, self-perception provides an additional inner perspective (cf. \cite{ulbrich2017a}).
Hence we propose the following definition:
\begin{quote}
	\small
	\textit{%
		Self-perception describes the process of generating knowledge about the internal state of the system by means of (model-based) data and information processing.
		Raw data for processing can be acquired by internal (proprioceptive) sensors or environment (exteroceptive) sensors.
		The models applied to infer and store knowledge about the internal state are part of the system's self-representation.}
\end{quote}

With these definitions, we can also establish the relation of self-perception and self-representation to self-awareness.

\subsubsection{Relation to Self-Awareness}
Following the proposition by \citeauthor{gregory2016} \cite{gregory2016} cited above, self-perception and self-representation are the fundamental properties which separate a self-aware system from a non-self-aware system.
Only by gathering knowledge about its current state (self-perception) and the required models to store the results (self-representation), the system can properly predict its actions and, more importantly, its future capabilities as demanded by \cite{gregory2016}.

	\section{Self-Perception \& Self-Representation for Safe Decision Making}
\label{sec:dec_making}
In our view, a comprehensive self-representation and self-perception is also a decisive factor for a vehicle automation system to execute safe actions at all (strategic, tactical, reactive) architectural levels.
In particular, we see it as a key element in planning safe behavior under uncertainty and by that as a key contribution to increased autonomy of vehicle systems.

To support this argument, we will shortly revisit our understanding of the ODD, as it establishes the system's general limits of operation and thus the general limits of the vehicle's action space (cf. \autoref{sec:action_space}).
Following this, we will argue why the uncertainty that is inherent to the formulation of ODD boundaries entails the need for monitoring.

\subsection[Aspects of the Operational Design Domain - Uncertain ODD Boundaries]{Aspects of the Operational Design Domain -- Uncertain ODD Boundaries\footnote{Section maybe extended in future revisions.}}
\label{sec:odd}
The ODD as a concept has gained a lot of traction for safety considerations for automated vehicles \cite{colwell2018, koopman2019, gyllenhammar2020}.
However, the definition of the ODD as
\begin{quote}
	\small
	\textit{%
"Operating conditions under which a given driving automation system or feature thereof is specifically designed to function, including, but not limited to, environmental, geographical, and time-of-day restrictions, and/or the requisite presence or absence of certain traffic or roadway
characteristics."} \cite[p.14]{sae2018}
\end{quote}
is rather broad.
The same is true, as \citet{koopman2019} note, for the ODD elements presented in \cite{nhtsa2017}.
While both references \cite{sae2018, nhtsa2017} make it clear that the ODD is more than just a geofencing mechanism, specifics are not given.

For this reason, \cite{koopman2019} present a more comprehensive list of factors which constitute the ODD, as well as a number of factors contributing to relevant Object and Event Detection and Response (OEDR) \cite{sae2018} strategies.
While we see no need to extend their lists here, we would like to stress their side-comment on the relation between the ODD and OEDR factors in which they state that "Specific events might not be applicable if no associated relevant objects are encompassed by the ODD" \cite[p. 2]{koopman2019}:
In other words, the boundaries of the ODD have direct impact on which objects and events the automated driving system needs to account for and are thus already by themselves a key source for requirements in all parts of the system.
In our view, this has two implications which concern the general formulation of ODD boundaries, and the complexity which is allowed within a given ODD, respectively.

A first important fact is that moving in public traffic and thus the decisions made by the vehicle automation system always come with an inherent risk.
This risk is caused by the fact that our traffic system basically constitutes an open world, which causes significant epistemic uncertainty \cite{nolte2018}: almost anything can happen at any time.
In consequence, this risk is directly reflected in uncertainty about the boundaries of the selected ODD when specifying the vehicle automation system.

The formulation of an ODD with corresponding OEDR strategies can now be seen as an attempt to convert the originally open world into an axiomatic closed world under a set of (hopefully) explicitly stated assumptions \cite{nolte2018, koopman2019a}.
By this, the influence of the risk that originates from the traffic system itself can be mitigated by choosing smaller ODDs (cf. $\mu$ODDs in \cite{koopman2019a}) and/or appropriate OEDR strategies \cite{nolte2018}.
However, the inherent risk can never be fully eliminated, as the assumptions about ODD boundaries and OEDR strategies remain subject to uncertainty and are hence potentially invalid.
-- Except, maybe, when choosing a maximally conservative design for situation prediction, decision making and trajectory generation which would compromise mobility (cf. \cite{nolte2018, koopman2019a}).

At the same time, increasing complexity of an ODD can at some point cause a combinatorial explosion of possible scenarios (cf. \cite{koopman2019}) which introduces additional uncertainty about incomplete system specification and validation, respectively.

\subsection{An Argument for Decision-Making Supported by Self-Representation and Self-Perception}
Given our understanding of the ODD as described above, we will now take a step back and provide a possible argument for a system implementation which does not rely on self-representation and -perception.
We will make a counterfactual argument assuming a \emph{fully robust} design approach.

When designing automated vehicles for safe behavior, an obvious option is to take a classic approach from safety engineering: Implementing a fully diversely redundant and thus robust and fail-operational automation system.
This would include fully redundant hardware platforms, software modules and algorithms, as well as sensors and actuators.
In case of sub-system failures (hence system-level errors for the automated vehicle as a whole), such a system would have little to no requirements to adapt its behavior decisions to these sub-system failures.
This design would make a propagation of sub-system failures beyond the overall system boundary extremely unlikely.
Thus the vehicle's overall performance would not be impaired.
Monitoring efforts could be reduced to classic approaches such as majority voting in order to detect when to switch to a redundant component.
A holistic system-level monitoring approach as discussed above would not be required.

An argument from this point of view could come to the conclusion, that the representations discussed in \autoref{sec:self-representation} would thus not be required, as well.
However, this would still neglect the influence of an ODD formulation as described in \autoref{sec:odd}:
When formulating ODD boundaries as a way to simplify the vehicle's operational space, this creates the need for monitoring if the vehicle is operating within these limits.
Uncertain ODD boundaries increase the need for monitoring in a way as this monitoring must be robust enough to account for this uncertainty:
The system must be able to assess how close the vehicle is operating to its ODD limits and how likely it is that the boundaries will be violated.

For this the system needs representations to predict its own behavior.
These representations must at least include dynamics models (cf. \autoref{fig:bdd}), and an idea of the ego vehicle’s shape.
In addition, the vehicle also needs to represent if it is still able to respond to objects and events (i.e. implement OEDR strategies) properly.
This can be seen as a representation of the by-design capabilities needed for operation in a defined ODD.

In summary, most aspects of self-representation mentioned in \autoref{sec:self-representation} remain important inputs to decision making, even when opting for a robust system implementation rather than opting for the system-level monitoring approach argued in \autoref{sec:self-representation}.
At least, this is true when designing a system so that it is able to respect its ODD at runtime, which is the only possibility to ensure safe ODD-conforming behavior.
This view is supported by recent publications which formulate the need to monitor the vehicle's ODD \cite{colwell2018, koopman2019, gyllenhammar2020}
\footnote{%
	While \cite{gyllenhammar2020} disregard self-representation due to a number of misinterpretations of the concepts presented in \cite{ulbrich2015}, they introduce the concept of "internal operating conditions", which "[...] are the conditions pertaining to the ADS itself and its user" \cite[p. 5]{gyllenhammar2020}. It remains unclear to us what "internal operating conditions" should reflect apart from the of the system's state and properties.%
}.
In a fully robust design approach, the only aspects of the self-representation which would lose their relevance is the representation of the architecture framework and with it the representation of non-functional requirements.
These could be spared due to the available redundant elements in all architecture views.

Additionally, we think a fully robust setup to be an unlikely design choice, mainly due to the cost related to the implementation of a fully redundant system, and physical limitations e.g. of sensor technology.
In consequence, we think that a self-aware design approach, which considers system-level performance monitoring as well as a coherent self-representation and self-perception for decision making, is a far more promising approach to implement safe vehicle behavior.

\subsection{Combining Internal and External Representations}
\label{sec:action_space}
According to \cite{ulbrich2017a}, the system makes decisions based on the assessment of a \emph{situation}.
I.e. the system interprets a \emph{scene} by selecting and augmenting the information in the scene with goal- and value-specific information which is relevant for the driving function.
This selection and augmentation is performed during situation assessment.
It can be implemented, amongst others, by filtering objects which have no chance of interacting with the vehicle, by predicting other traffic participants' behavior, by combining the scene with the automated vehicle's desired route, or by selecting relevant traffic rules.
This, in turn, means that an evaluation or judgment of collected knowledge about the environment is not performed during \emph{scene} creation.

The same is true for the internal representations of the vehicle:
While its internal models are applied to infer and store knowledge e.g. with respect to the system's capabilities during scene creation, no assessment is performed, whether the system's capabilities are sufficient to perform a certain maneuver.
This is only performed during situation assessment, when the system's (mission) goals (and values) are known.
Here, knowledge obtained through self-perception and stored in the self-representation can be used to influence decision making.
In analogy to the concept of $\mu$ODDs presented in \cite{koopman2019a} and the concept of restricted operational domains (RODs) presented by \cite{colwell2018}, self-representation and self-perception are, in our view, key to ensure that system actions keep the system inside a safe operational domain.

\begin{figure}
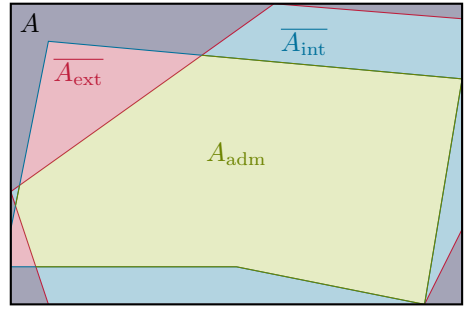

	\centering
	\includeimg{action_space.tikz}
	\caption{Restrictions of the vehicle's action space in analogy to the presentation of the Restricted Operational Domain in \cite{colwell2018} - black: total action space (resulting from the design-time system capabilities which allow operation in the system's ODD), red: restrictions imposed by elements in the vehicle's environment, blue: runtime-restrictions imposed by self-perception (due to degraded capabilities), green: admissible action space.}
	\vspace{-1em}
	\label{fig:sets}
\end{figure}

For a more formal description, we borrow the notation of a Markov decision process:
Given a state space $S$ and a (possibly discrete) action space $A$, consider the system being in a state $s \in S$.
The entirety of $A$ is the set of actions which is based on all system capabilities required to operate in the system's target ODD (which would be an open world for SAE Level 5 systems).
The general challenge when making decisions is to ensure that an action $a \in A$ is only taken if the resulting state $s' \in S$ is inside a subset of $S$ which corresponds to a safe operational domain.

Hence, the results of environment- and self-perception both restrict the vehicle's action space:
Actions leading to an unacceptable risk emerging from the vehicle's behavior must be avoided.
Restrictions from environment perception e.g. should avoid actions which cause unacceptable collision probabilities.
Restrictions from self-perception e.g. should avoid actions for which the system has insufficient capabilities.
An example for the latter restrictions would be to avoid lane changes if the vehicle has insufficient sensor coverage to its rear.

With these assumptions, the action space $A$ can be partitioned into two subsets (cf. \autoref{fig:sets}):
Subset $\overline{A_\mathrm{ext}} \subseteq A$ is inadmissible due to restrictions from environment perception and prediction.
Subset $\overline{A_\mathrm{int}} \subseteq A$ is inadmissible due to restrictions from self-perception and the anticipated internal system states.

Hence the admissible set of actions $A_\mathrm{adm}$ which keeps the vehicle in a safe operational domain is the cross section $A_\mathrm{adm} = A_\mathrm{env} \cap A_\mathrm{int}$ (note that in general $A_\mathrm{env} \cap A_\mathrm{int} \neq \emptyset$).

In the following section we will give an example how self- and environment representation actually restrict the admissible action space of an automated vehicle.
	\section{Example Scenario}
\label{sec:example}
\begin{figure}[t!]
	\includeimg[trim=0 20 0 0, clip, width=.99\columnwidth]{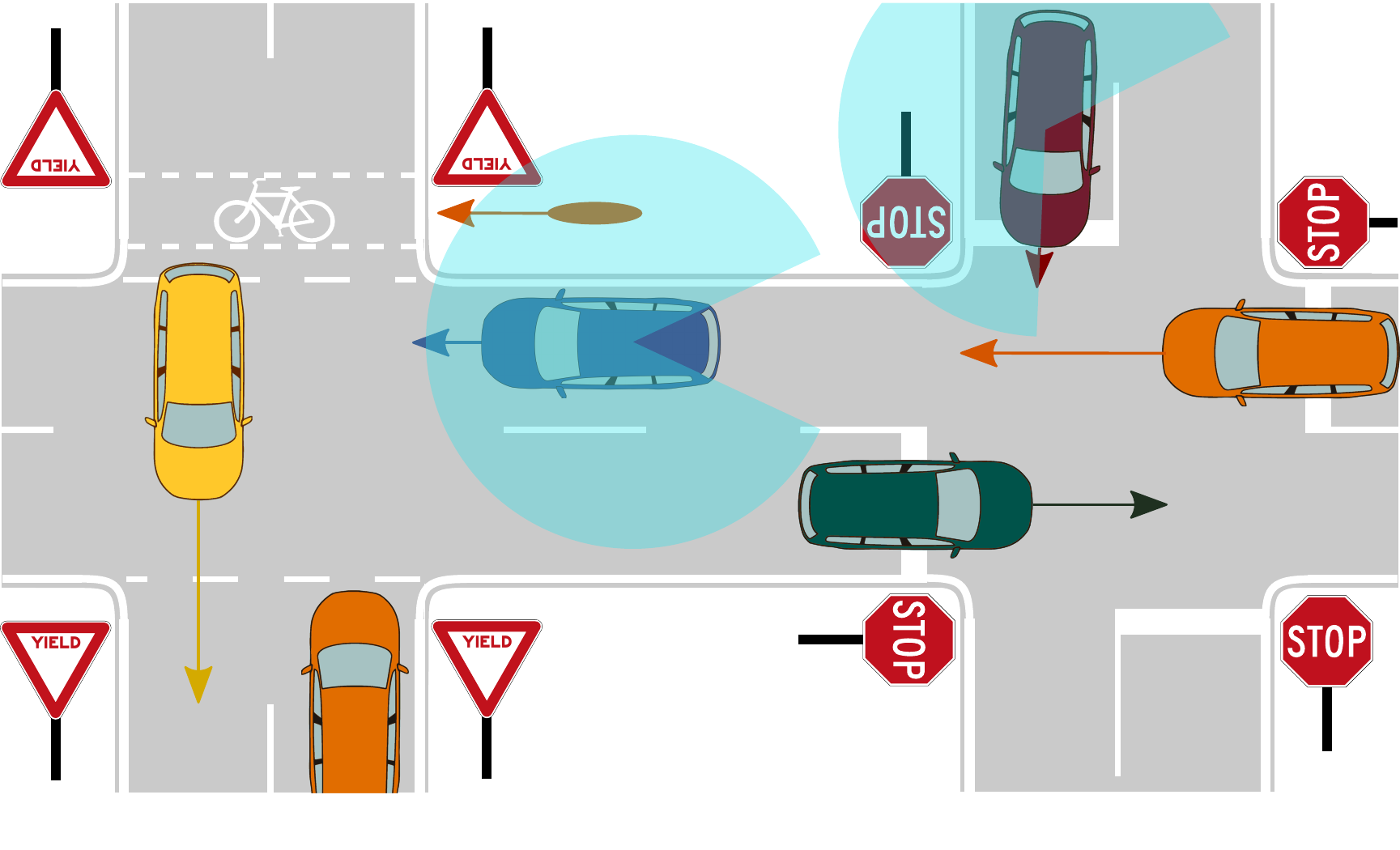}
	\caption[Caption]{(Objective) initial scene of example scenario: Scene from an observer's point of view. Arrows indicate objects' current velocities, orange ellipse represents a cyclist. Light blue areas depict sensor coverage of two automated vehicles (only showing angular range; radial range is assumed to cover the complete scene). Illustration extended\footnotemark{} from \cite[p. 986, Fig. 6]{ulbrich2015}.}
	\vspace{-1em}
	\label{fig:scene}
\end{figure}

To illustrate the aspects of self-perception and self-representation in the context of scene modeling and situation assessment further, we will refer to an example scenario.
The start scene of the scenario is shown in \autoref{fig:scene}.
No viewpoint is chosen, i.e. the objective scene is displayed.
The vehicles in the center (blue) and top right (dark red) are automated.
A cyclist (orange) is riding on a separate bike lane in parallel to the blue automated vehicle.
No inactive observers are present as elements in the scene.
\addtocounter{footnote}{-1}%
\footnotetext{Modified with the author's permission. Original graphic \copyright 2015 IEEE}%

Note that for an objective scene description by an omniscient observer as shown in Fig. 3, according to [9], both vehicle’s self-representations would be part of the scene description. 
Considering e.g. a simulation environment, the simulation system can be seen as an omniscient actor in the scene would have knowledge about the stored self models of both vehicles.

For the sake of providing a visual example how self-representation impacts the scene and situation, we assume that both automated vehicles do not have full sensor coverage:
While the blue vehicle is blind to its rear, the red vehicle cannot perceive anything to its left.
Apart from this, we assume fully intact mechanics, actuators, and computing hardware.
All algorithms are assumed to perform as intended.

Models in the self-representations of both vehicles, in particular the capability representations, store this information in terms of the vehicles' remaining field of view and in terms of the stored capability metrics.
In order to assess the current situation, the goals and values of the blue and red vehicle must be considered respectively.
For the given scenario, we assume that both vehicles have mission objectives which result in the goal to pass the intersection going straight.

\begin{figure}[t!]
	\includeimg[trim=0 20 0 0, clip, width=.99\columnwidth]{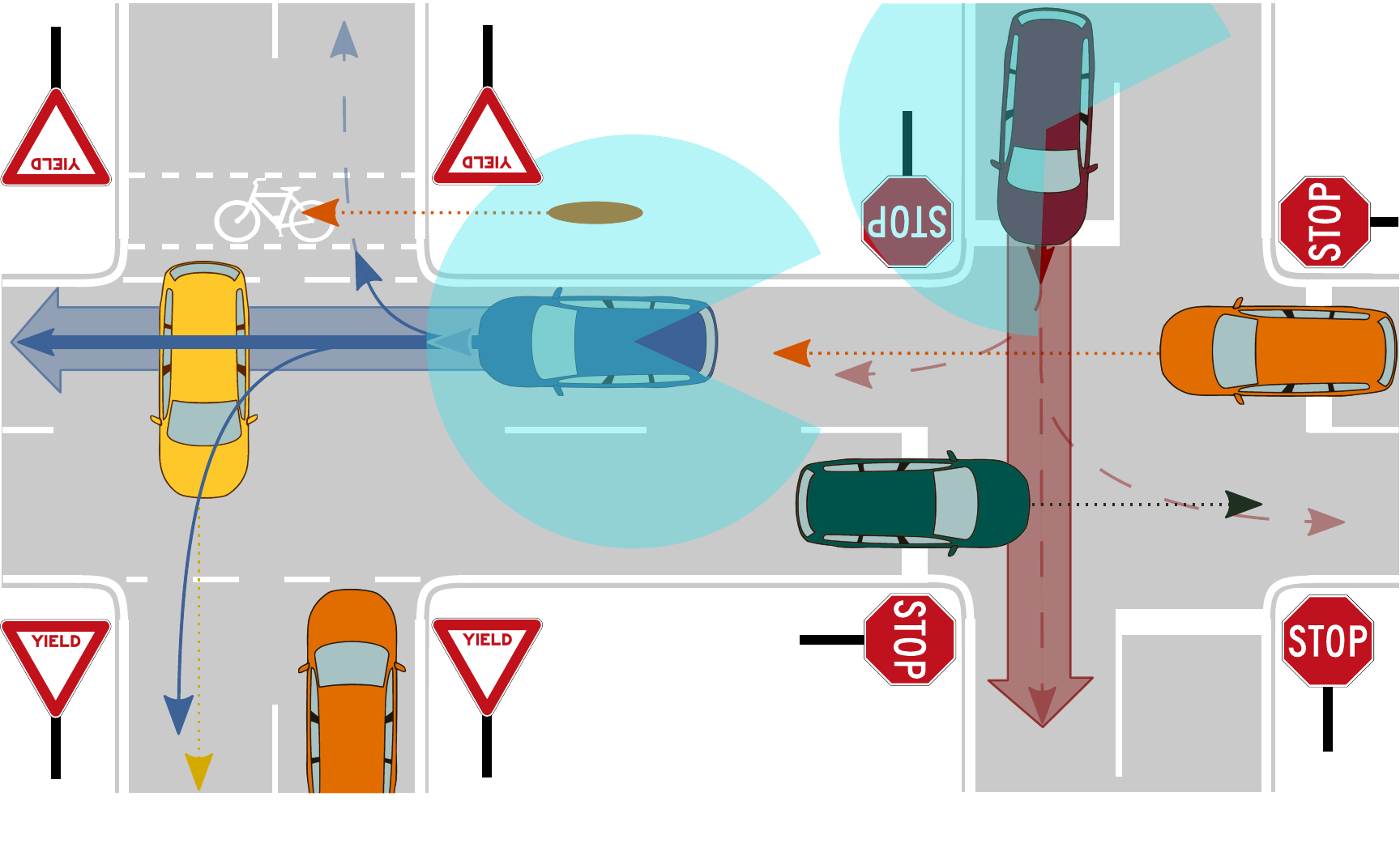}
	\caption{(Ground-Truth) Situation derived from the scene shown in \autoref{fig:scene}: Dotted arrows indicate predicted trajectories of objects, transparent block arrows indicate mission objectives. Transparent dashed/opaque solid arrows indicate action spaces without/with consideration of self-representation, predictions and mission objectives. Illustration extended$^\thefootnote$ from \cite[p. 986, Fig. 6]{ulbrich2015}.}
	\vspace{-1em}
	\label{fig:situation}
\end{figure}

In the following, we will discuss the impact of both vehicle's self-representations separately at the example of \autoref{fig:situation}.
For all non-automated traffic participants, we assume straight, constant-speed trajectories.

For the \emph{blue} vehicle, its degraded sensor performance has little impact on the vehicle's mission.
It has full view of all arms of the intersection and is also able to perceive the cyclist riding in the bike lane.
It can perceive the red and green vehicles as well, while the orange vehicle is not part of the blue vehicle's environment representation.

If we consider the vehicle's action space as a set of maneuvers, possible actions the vehicle could take by design are: stopping, turning right, turning left, or passing the intersection going straight (blue arrows in \autoref{fig:situation}).
In this situation, restrictions of the action space from the vehicle's self-representation only concern stopping maneuvers:
As the vehicle cannot determine whether another vehicle is following closely, harsh braking should not be performed.
Action space restrictions from the environment representation only concern a possible right turn.
According to the prediction of the cyclist, there is a chance of collision when the vehicles' actions correspond to the transparent dashed arrow in \autoref{fig:situation}.
The resulting admissible action space hence contains maneuvers which lead to stopping slowly, letting the cyclist pass, turning left while staying behind the yellow vehicle, and crossing the intersection straight.

The vehicle can also use the models in its self-representation to simulate its own behavior (e.g. predicting its motions along the thin arrows in \autoref{fig:situation}) to evaluate its options and assess the most beneficial action with respect to its mission.
As can be seen in \autoref{fig:situation}, the blue vehicle can fulfill its mission without major restrictions:
The most beneficial action with respect to the vehicle's mission goal would be to go straight.

In contrast to this, the \emph{red} vehicle is severely impacted by its remaining sensing capability.
As it cannot perceive the right arm (from the topview) of the intersection, the only admissible action in this situation is stopping.
The vehicle is lacking the required performance to fulfill its mission.
	\section{Conclusion}%
\label{sec:conclusion}%
In this paper we presented an overview about functional monitoring approaches for automated vehicles.
In this context, we motivated the need for a coherent representation of internal models, i.e. knowledge about the system's current internal state.
We gave definitions for the terms \emph{self-perception} and \emph{self-representation} for automated vehicles consistent with existing definitions for the terms \emph{scene} and \emph{situation}, and provided a taxonomy for the contents of the self-representation.

We provided an example scenario to illustrate the impact of self-representation for situation assessment and safe decision making.
What becomes clear from the given scenario is that an automated vehicle's self-representation is highly relevant for a coherent situation assessment.
At the same time, only by providing situational context, such as the vehicle's goals, the knowledge stored in the vehicle's self-representation can be applied properly for decision making.

This is, of course, true of all elements in the vehicle's self-representation (cf. \autoref{fig:bdd}), not only for the sensing capabilities which we used as an example here.

In summary, a \emph{sufficiently detailed self-representation} must contain explicit representations of all relevant system-internal aspects for planning and executing safe vehicle behavior in a target ODD.

\section*{Acknowledgement}
We would like to thank Till Menzel for valuable input on the definitions of scene and situation and Simon Ulbrich for providing permission to modify his original graphics \cite{ulbrich2015}.
	\renewcommand*{\bibfont}{\footnotesize} 
	\printbibliography 
	\begin{IEEEbiography}[{%
			\includegraphics[width=1in,height=1.25in,clip,keepaspectratio]{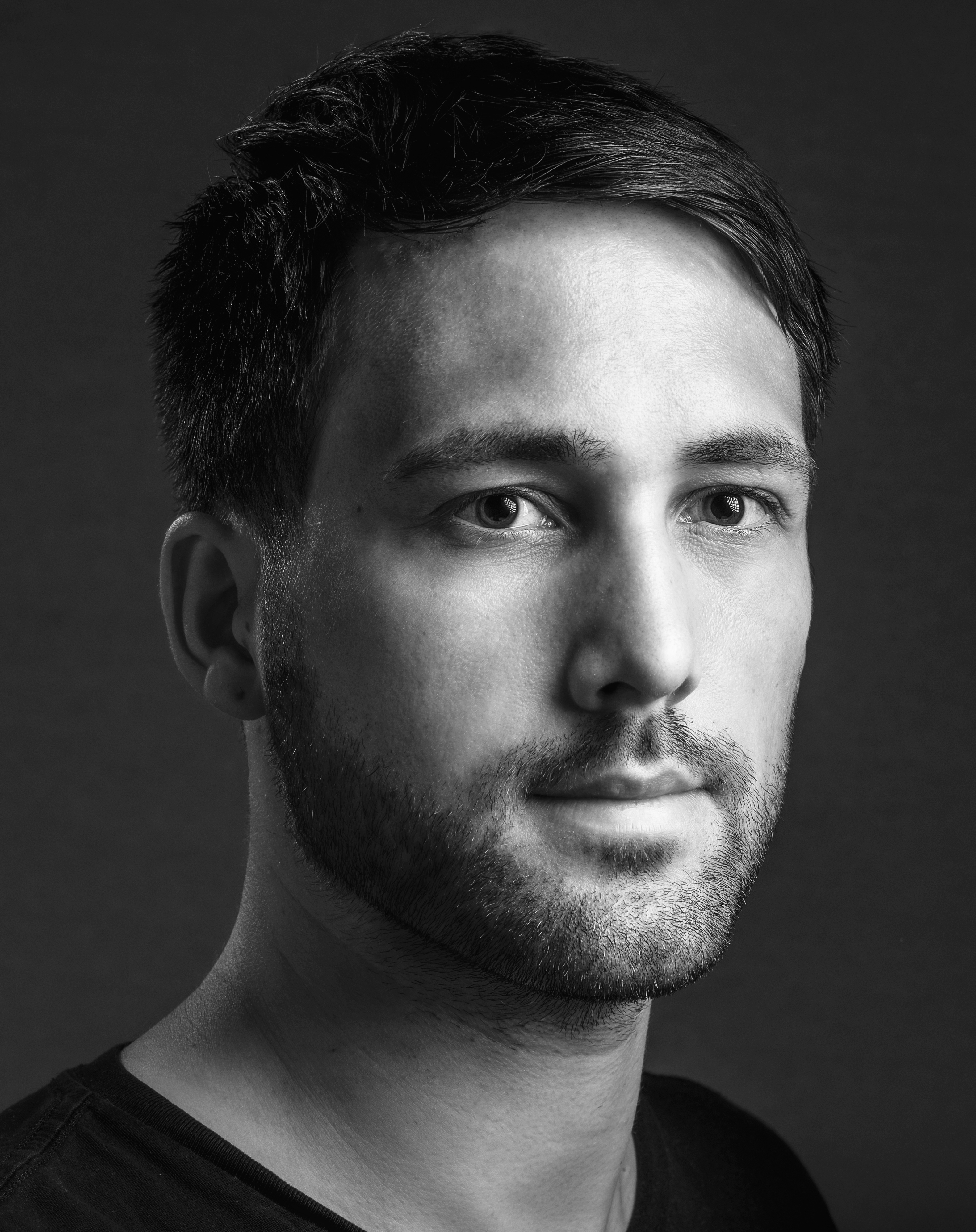}%
		}]{Marcus Nolte}
		works as a research assistant at the Institute of Control Engineering at TU Braunschweig since 2014 and is currently pursuing his PhD. 
		He received his Master of Science in Electrical Engineering	from TU Braunschweig. 
		His main research interest is self- and risk-aware and motion planning for automated vehicles.
	\end{IEEEbiography}

	\begin{IEEEbiography}[{%
		\includegraphics[width=1in,height=1.25in,clip,keepaspectratio]{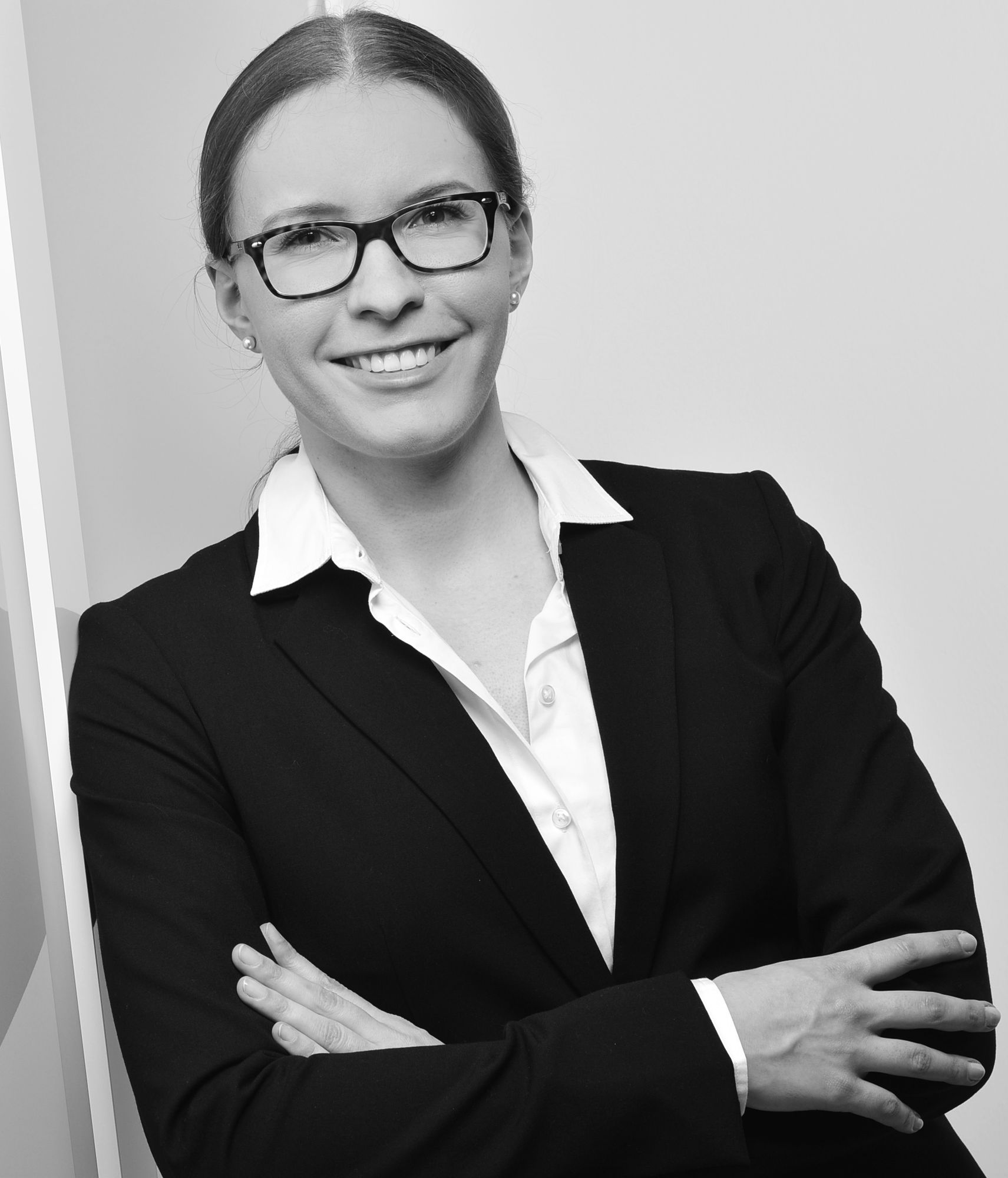}%
	}]{Inga Jatzkowski}
		works as a research assistant at the Institute of Control Engineering at TU Braunschweig since 2016 and is currently pursuing her PhD. 
		She holds a Master of Science Degree in Navigation and Field Robotics from Leibniz University Hannover. 
		Her main research topics are self-awareness and the development of self-perception for automated vehicles.
	\end{IEEEbiography}
	\begin{IEEEbiography}[{%
		\includegraphics[width=1in,height=1.25in,clip,keepaspectratio]{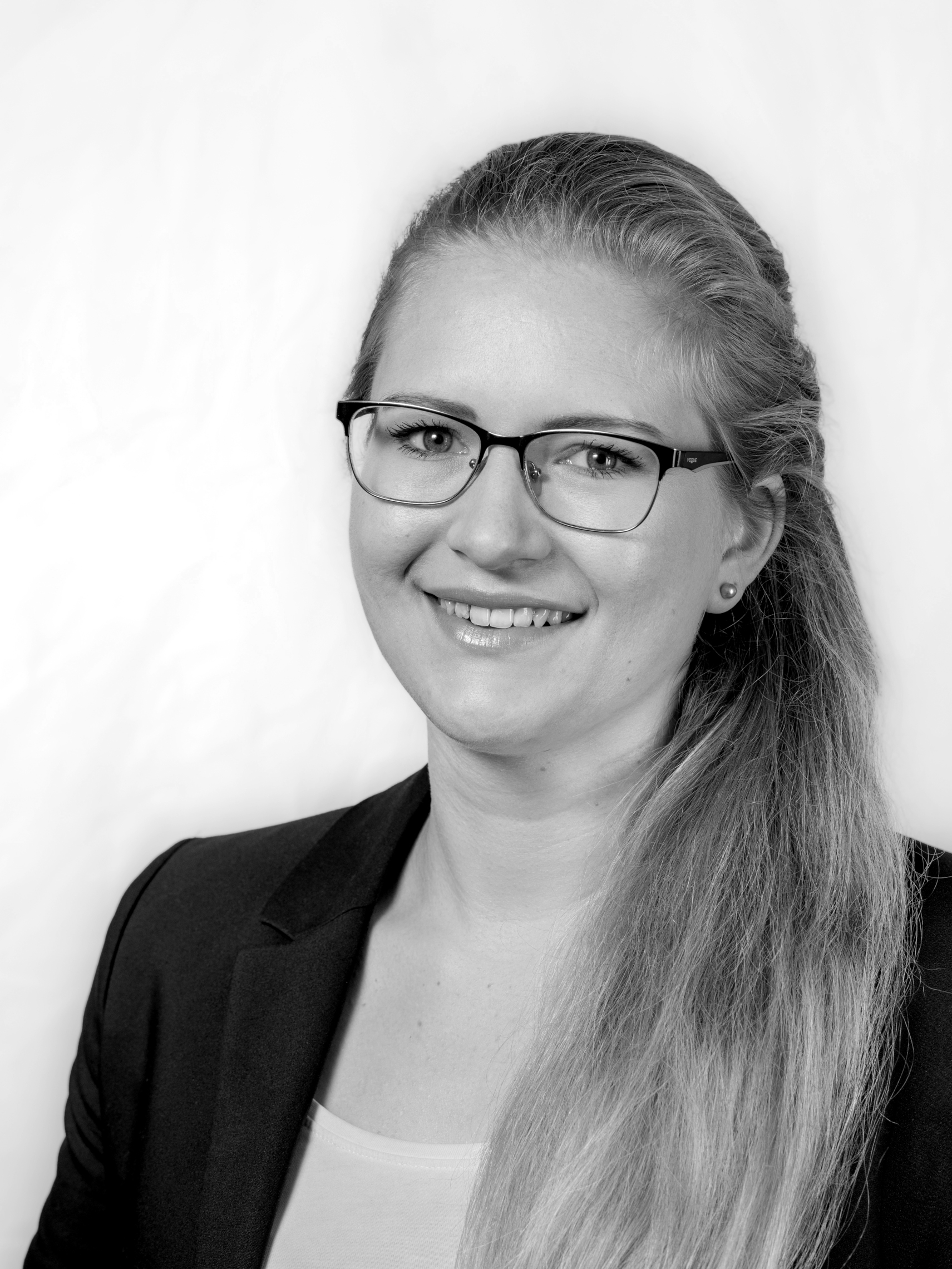}%
	}]{Susanne Ernst}
		works as a research assistant at the Institute of Control Engineering at TU Braunschweig since 2015.
		She received her Master of Science in Mechanical Engineering from TU Braunschweig. 
		Her main research field is risk assessment for decision making of automated vehicles.
	\end{IEEEbiography}

	\begin{IEEEbiography}[{%
		\includegraphics[width=1in,height=1.25in,clip,keepaspectratio]{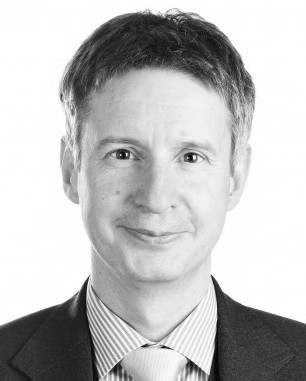}%
	}]{Markus Maurer}
		holds the chair for Vehicle Electronics at TU Braunschweig since 2008. 
		His main research interests include autonomous road vehicles, driver assistance systems, and automotive systems engineering. 
		From 2000 to 2007 he was active in the development of driver assistance systems at Audi AG.
	\end{IEEEbiography}
\vfill
\end{document}